\documentstyle[aps,amsfonts]{revtex}

\expandafter\ifx\csname mathbb\endcsname\relax
 \let\mathbb=\Bbb
 \let\mathcal=\cal
\fi

\newtheorem{lemma}{Lemma}
\newtheorem{proposition}{Proposition}
\newtheorem{definition}{Definition}[section]

\def\mem{\,\epsilon\,}

\def\reals{{\mathbb R}}

\def\boga{\alpha_{lk}}
\def\bogb{\beta_{lk}}
\def\bogabar{\bar{\alpha}_{lk}}
\def\bogbbar{\bar{\beta}_{lk}}

\def\ckgproduct#1#2{ (#1,#2)_{KG} }

\def\thetafm{\Theta_{y}}
\def\deltafm{\Delta_{y}}
\def\rthetafm{\Theta_{\lambda}}
\def\rdeltafm{\Delta_{\lambda}}

\def\data{ {{\phi} \choose {\pi}} }

\def\bar#1{\overline{#1}}

\def\vac{|0\rangle}
\def\rvac{|0^{R}\rangle}

\def\uk#1{U_{\vec{#1}}}
\def\ukstar#1{\overline{U}_{\vec{#1}}}

\begin{document}


\title{
\hfill{\normalsize 10 April 2000}
\\[3mm]
{\bf DISTRIBUTIONAL MODES FOR SCALAR FIELD QUANTIZATION}
}

\author{Alfonso F. Agnew}
\address{
Department of Mathematics,
Southern Methodist University, Dallas, TX  75275, USA \\
{\tt aagnew{\rm @}mail.smu.edu}
}

\author{Tevian Dray}
\address{
Department of Mathematics,
Oregon State University, Corvallis, OR  97331, USA \\
{\tt tevian{\rm @}math.orst.edu}
}

\maketitle

\begin{abstract}
We propose a mode-sum formalism for the quantization of the scalar field based
on distributional modes, which are naturally associated with a slight
modification of the standard plane-wave modes.  We show that this formalism
leads to the standard Rindler temperature result, and that these modes can be
canonically defined on any Cauchy surface.
\end{abstract}

\section{INTRODUCTION}

Manogue et al.\ \cite{MDCop} considered the quantization of the scalar field
on the trousers spacetime, showing that the natural plane-wave bases are not
complete in this setting.  Specifically, they pointed out the existence of
distributional modes which vanish identically on one region, and which
correspond to initial data at the singular point joining the regions.  They
concluded by speculating on the possibility of quantizing this extra degree of
freedom.  It is then natural to consider quantizing the scalar field {\it
entirely\/} in terms of such distributional modes; this work represents an
attempt to do just that.

There is nothing new in using distributions to describe the scalar field.
After all, the Feynman propagator is a distribution, as are the various other
2-point ``functions'' used, among other things, to define the notion of
positive-frequency.  What is different about the approach outlined here is
that we choose to treat certain distributions as basis modes, which we use to
expand the field for the purpose of second quantization.

Our distributional modes have {\it position\/} space indices, and in this
sense are dual to the usual {\it momentum\/} space description in Minkowski
space.  But the duality is not precise in that the natural distributional
modes correspond not to the standard plane-wave modes, but rather to what we
call {\it pseudo\/} plane waves.  We therefore also investigate these latter
modes.  We show that they are a member of an infinite family of distinct mode
sets that can be used in place of the standard plane waves to derive the
Rindler temperature.

To motivate our general discussion, we first consider 2-dimensional Minkowski
space, explicitly constructing a family of distributional modes and discussing
both their relationship to standard plane wave modes and their use in
computing the Rindler temperature.  This is the case that would apply to the
trousers spacetime, after taking into account the periodicity the trousers
impose.  We then use properties of the fundamental solution of the
Klein-Gordon equation to establish the existence and uniqueness of canonical
distributional modes on any globally hyperbolic spacetime.

We review the standard description of the scalar field in
Section~\ref{Background}, introduce distributional modes in 2-dimensional
Minkowski space in Section~\ref{2d}, and discuss the Rindler temperature
calculation in Section~\ref{Rindler}.  In Section~\ref{Cauchy} we then show
how to define canonical distributional modes on any Cauchy surface in any
globally hyperbolic spacetime, and in Section~\ref{Discussion} we discuss our
results.

\section{BACKGROUND}
\label{Background}

In this section we review some basic properties of the quantum scalar field in
Minkowski space, such as mode decomposition, positive-negative frequency
decomposition, and the determination of the Fock space of states by the
characterization of the vacuum.

\goodbreak
Consider a globally hyperbolic spacetime $({\cal{M}}, g)$.  The classical
relativistic scalar field $\Phi$ on $\cal M$ satisfies the Klein-Gordon
equation
\begin{equation}
 (\Box - m^2)\,\Phi = 0,\
 \label{eq:generalwaveeqn}
\end{equation}
where $\Box := g^{ab}\nabla_{a}\nabla_{b}$ and our signature is
($-$ $+$ $+$ $+$).
\footnote{A more general form of the Klein-Gordon  equation includes a term
proportional to the scalar curvature.  In setting this term equal to zero, we
consider the standard case of minimal coupling \cite{Fulling89}.  For more on
the deceptively difficult issue of nonzero coupling, see
\cite{Fischerthesis,Fischer}.}
The Klein-Gordon form for complex-valued solutions of
(\ref{eq:generalwaveeqn}) with compact Cauchy data is
\begin{equation}
 \ckgproduct{\Phi}{\Psi} := -i\int_{\Sigma}\, \Big{(} (\nabla_{\hat n}
\overline{\Phi})
 \Psi - \overline{\Phi} \nabla_{\hat n}
  \Psi \Big{)} \, d\Sigma,
 \label{eq:solsympdef}
\end{equation}
where $d\Sigma$ is the canonical volume element on the Cauchy surface
$\Sigma$, and ${\hat n}$ is the future pointing unit normal to $\Sigma$.  This
form is independent of the choice of $\Sigma$, and is more commonly recognized
as the inner product for the set of ``positive frequency'' solutions.

The standard theory in Minkowski space uses the Fourier transform to express
the field $\Phi$ as a mode-sum over plane waves,
\begin{equation}
\Phi(t,\vec{x}) = \int_{\vec{k}}\; \Big{(}a(\vec{k}) U_{\vec{k}} +
\bar{a}(\vec{k})\bar{U}_{\vec{k}}\Big{)}\, d\vec{k},
\end{equation}
with
\begin{eqnarray}
U_{\vec{k}}
&=& N_{\vec{k}} e^{-i\omega_{\vec{k}} t + i\vec{k}\cdot\vec{x}} \\
\bar{U}_{\vec{k}}
&=& N_{\vec{k}} e^{i\omega_{\vec{k}} t - i\vec{k}\cdot \vec{x}},
\end{eqnarray}
and where $\omega_{\vec{k}} = \sqrt{\vec{k}^{2} + m^{2}}.$ The
$\vec{k}$-dependent normalization constants are given by
$N_{\vec{k}} = {\frac{1}{\sqrt{4\pi\omega_{\vec{k}}}}}$.
The normalization of the plane wave modes is given by
\begin{eqnarray}
\ckgproduct{U_{\vec{k}}}{U_{\vec{\tilde{k}}}}
& = & \delta(\vec{k} - \vec{\tilde{k}} )  
\label{pwonor1}     \\
\ckgproduct{\bar{U}_{\vec{k}}}{\bar{U}_{\vec{\tilde{k}}}}
& = & -\delta(\vec{k} - \vec{\tilde{k}}) 
 \label{pwonor2}   \\
\ckgproduct{U_{\vec{k}}}{\bar{U}_{\vec{\tilde{k}}}}
& = & 0.
\label{pwonor3}
\end{eqnarray}
A positive-negative frequency splitting is made by choosing the
$\{U_{\vec{k}}\}$ as a basis for positive frequency solutions.  The positive
frequency part of a (real-valued) solution is then given by
\begin{equation}
 \Phi^{+} := \int_{\vec{k}} a(\vec{k}) e^{i\vec{k} \cdot \vec{x} -
 i\omega_{\vec{k}}t} {d^{3}k \over
 \sqrt{2\omega_{\vec{k}}(2\pi)^{3}}} =
 \overline{\Phi^{-}}.
 \label{eq:positivefrequencypart}
\end{equation}
In quantization, the expansion coefficients $a(\vec{k})$,
$\overline{a}(\vec{k})$ become operators $a(\vec{k})$, $a^\dagger(\vec{k})$
satisfying the appropriate commutation relations.  The vacuum state is then
characterized by the requirement
\begin{equation}
 a(\vec{k})\vac = 0 ,\; \forall \; \vec{k}.
 \label{vacuumcondition}
\end{equation}
The final Hilbert space of states (Fock space) is generated by elements of 
the form \cite{BirrellandDavies} 

\begin{equation}
a^\dagger({\vec{k}})\vac.
\end{equation}

In a curved spacetime, there is a standard mode-sum formalism generalizing the
Minkowski space procedure outlined above \cite{BirrellandDavies}.  In fact,
the only part of the above procedure that is specific to Minkowski spacetime
is the choice of plane wave solutions for the mode-sum.  Since there do not
exist plane waves solutions to the Klein-Gordon equation in a general curved
spacetime, one instead chooses a set $\{U_{\vec{k}}, \bar{U}_{\vec{k}}\}$ of
complex solutions satisfying the same orthonormality conditions
(\ref{pwonor1}) -- (\ref{pwonor3}).  Furthermore, the solutions must form a
complete set, in that the real fields must be expandable as a mode-sum over
the $\{U_{\vec{k}}, \bar{U}_{\vec{k}}\}$.  Of course, finding such a set of
solutions, and proving that they have the required properties to use in a
mode-sum expansion, can be a very difficult problem.  As we shall see, the
distributional modes formalism is able to overcome these difficulties.


\section{Two-Dimensional Distributional Modes}
\label{2d}

\subsection
 {The Distributional Mode Expansion and Standard Quantum Field Theory}

The distributional modes first arose in the context of the $2$-dimensional
flat, but topologically nontrivial, trousers spacetime \cite{MDCop}.  We
suggest that these modes (or an appropriate generalization) may similarly
prove useful when considering quantization on other spacetimes that do not
possess such nice properties as global hyperbolicity.
\footnote{Indeed, there has been an increasing interest in the literature with
these issues \cite{KayRadWald,Kaynongh,Hawking92}, especially  in connection
with Hawking's ``Chronology Protection Conjecture''.}
As a first step in this development, we begin with the case of two-dimensional
Minkowski space ${\mathbb{M}}^2$.  In this case, the distributional modes
retain the same form as in \cite{MDCop}.  Also, the well established plane
wave-based theory in Minkowski space can then be used to guide the
generalization of the distributional modes theory to other spacetimes.

The modes considered by Manogue et al.\ satisfy the wave equation
(\ref{eq:generalwaveeqn}) in the distributional sense and are of the form
\begin{eqnarray}
\thetafm (t,x) & = &
 \frac{1}{2}[\theta(x - y + t) - \theta(x - y - t)] \label{2dtheta}\\
\deltafm (t,x) & = &
 \frac{1}{2}[\delta(x - y + t) + \delta(x - y - t)], \label{2ddelta}
\end{eqnarray}
where $y \mem {\mathbb{R}} \cong \Sigma := \{(0,y)\mem {\cal M}\, \}$.

We consider a formalism for the canonical quantization of the scalar field
based on this set of modes, which differ from the usual plane wave modes.  In
particular, the new modes are indexed by {\it position} on the surface $t=0$,
as opposed to {\it momentum}, are real instead of complex, and satisfy:
\begin{eqnarray}
\ckgproduct{\thetafm}{\Theta_{\tilde{y}}} & = & 0 \\
\ckgproduct{\deltafm}{\Delta_{\tilde{y}}} & = & 0 \\
\ckgproduct{\thetafm}{\Delta_{\tilde{y}}} & = & -i\delta(y - \tilde{y}).
\end{eqnarray}
The first step in such a formalism is to expand the field in terms of a set of
basis modes, so assume
\footnote{This is a nontrivial assumption in general! See also \cite{DM}.}
that the field can be expanded uniquely in terms of either the plane wave
modes $\{U_{k}, \overline{U}_{k}\}$ or the distributional modes
$\{\thetafm,\deltafm\}$:
\begin{eqnarray}
\Phi(x,t) & = &
 \int_{k}\; \Big{(}a(k) U_{k} + \bar{a}(k) \bar{U}_{k}\Big{)}\, dk
 \label{minkopwexp}  \\
\Phi(x,t) & = &
 \int_{y} \; \Big{(}\phi(y) \deltafm + \pi(y) \thetafm\Big{)}\, dy.
 \label{minkofmexp}
\end{eqnarray} 
The relationship between the expansion coefficients may then be determined by
the Bogolubov coefficients
\begin{eqnarray}
a(k) & = & (U_{k},\Phi) \nonumber  \\
& = &  \int_{y} dy \Big{(}\phi (y) (U_{k },\deltafm) +
                 \pi (y) ( U_{k} ,\thetafm )\Big{)}  \\
& = & \int_{y} dy \Big{[} \pi (y)
 \Big{(} \frac{i}{2\sqrt{\pi \omega_{k}}} e^{-iky} \Big{)}
 + \phi (y) \Big{(}{1 \over 2}
 \sqrt{ {\omega_{k} \over \pi} } e^{-iky} \Big{)} \Big{]},
\end{eqnarray}
which can also be written in terms of a Fourier transform as
\begin{equation}
a(k)  =   \sqrt{\frac{\omega_{k}}{2}} \,\hat{\phi}(k)
 + i \,\sqrt{\frac{1}{2\omega_{k}}} \,\hat{\pi}(k).
 \label{akexpansion}
\end{equation}
Similarly, we find
\begin{equation}
\bar{a}(-k) = \sqrt{\frac{\omega_{k}}{2}} \,\hat{\phi}(k)
  - i \,\sqrt{\frac{1}{2\omega_{k}}} e\,\hat{\pi} (k).
\end{equation}
For fixed $k$, $a(k)$ is simply the annihilation operator associated with a
simple harmonic oscillator of frequency $k$ with phase space variables
$\hat{\phi}$ and $\hat{\pi}$!  Thus, as the plane wave expansion of the field
reveals the field as an infinite collection of harmonic oscillators indexed by
their fixed frequency, the distributional mode expansion reveals the field as
an infinite collection of Fourier transformed harmonic oscillators indexed by
their fixed {\it position\/} on the $t=0$ surface.  Notice that there is
Heisenberg duality at work here: the plane wave oscillators have definite
frequencies, and so are completely nonlocal, whereas the distributional mode
``oscillators'' are localized, and have indefinite frequency.  In terms of the
plane wave coefficients, the distributional mode coefficients are
\begin{eqnarray}
\phi(y)
& = & i\ckgproduct{\thetafm}{\Phi}  \\
& = &  i\int_{\reals}\; \Big{(}a(k) \ckgproduct{\thetafm}{U_{k}} +
   \bar{a}(k) \ckgproduct{\thetafm}{\bar{U}_{k}}\Big{)} dk \\
& = & \int_{\reals} N_{k}e^{iky}a(k) + N_{k} e^{-iky} \bar{a}(k)\; dk \\
& = & \Big{[}
 \widehat{\widehat{\left(\frac{a}{\sqrt{2\omega_{k}}}\right)}} (y) - 
 \widehat{\widehat{\left(\frac{\bar{a}}{\sqrt{2\omega_{k}}}\right)}}(-y)
 \Big{]}, 
\end{eqnarray}
where $\hat{\hat{f}\:\,}$ represents the inverse Fourier transform of $f$.
Similarly,
\begin{eqnarray}
\pi(y) & = &  -i\ckgproduct{\deltafm}{\Phi} \\
& = &  -i  \Big{(}
\widehat{\widehat{\left(\sqrt{2\omega_{k}}a\right)}} \, (y) - 
\widehat{\widehat{\left(\sqrt{2\omega_{k}}\bar{a}\right) }} \,(-y)\Big{)}.
\end{eqnarray}

The commutation relations for the operator coefficients $\phi(y)$ and $\pi(y)$
may be computed from the standard CCR imposed on the operator coefficients
$a(k)$ and $a^{\dagger}(k)$. The result is:
\begin{equation}
 [\phi(y),\pi(\tilde{y})] = i\delta(y - \tilde{y}),
\label{dmccr}
\end{equation}
with all other commutators vanishing which, of course, is just the equal time
CCR for the field operator and the momentum operator.  This is not a
coincidence and is the reason for denoting the distributional mode
coefficients as we have: $\phi(y) = \Phi(0,y),$ $\pi(y) = \partial_{t}
\Phi(0,y)$.  This will be addressed more fully in Section~\ref{Cauchy}, where
the general distributional mode formalism is discussed, but we note that this
validates our assumption that the field may be uniquely expanded in terms of
distributional modes.

The goal is to have a mode-sum formalism that views the distributional modes
and their operator coefficients as fundamental, replacing the use of plane
waves.  Furthermore, we want the formalism to reproduce standard flat space
quantum field theory.  Thus, we take the decomposition (\ref{minkofmexp}) as
our starting point, and impose the canonical commutation relations
(\ref{dmccr}) on the coefficients.  The Fock space of states is generated by
the vacuum, defined by (cf.~(\ref{vacuumcondition}))
\begin{equation}
 \Big{(}\sqrt{\frac{\omega_{k}}{2}}
 \hat{\phi}(k) + i \sqrt{\frac{1}{2\omega_{k}}} \hat{\pi}(k)\Big{)}
 \vac = 0, \; \; \; \; \forall k \mem \reals.
 \label{dmvaccondition}
\end{equation}
Equation (\ref{dmvaccondition}) defines the same vacuum as the plane waves,
but in terms of a new basis.

\subsection{Position Fock Space}

One drawback of the previous construction is the definition of the vacuum via
(\ref{dmvaccondition}), which requires a Fourier transform in the variable
$y$, which destroys the association between distributional modes and
individual points on a Cauchy surface.  This association is where much of the
potential utility of the distributional modes lies.  Further, problems can
occur with the Fourier transform when working in curved spaces.  However,
there is an alternative description of the Fock space that does not require a
Fourier transform of the (distributional mode) expansion coefficients $\phi$
and $\pi$, as in (\ref{dmvaccondition}), but only a positive-negative
frequency splitting of the field.  The vacuum condition
(\ref{vacuumcondition}) is equivalent to the condition that the vacuum is
annihilated by the positive frequency part of the field, as defined by
(\ref{eq:positivefrequencypart}):
\begin{equation}
 \Phi^{+}(t,x)\vac = 0 \; \; \; \; \forall x \mem \reals.
 \label{altvaccondition}
\end{equation}
Furthermore, as discussed in \cite{Roman}, if we denote the one-particle state
of momentum $k$ by
\begin{equation}
|1_{k}\rangle = a^{\dagger}(k)\vac,
\end{equation}
then the function $f_{k}(t,x) := \langle0|\Phi(t,x)|1_{k}\rangle$ is the
positive energy wave function in the configuration space version of the
nonquantized Klein-Gordon equation.  In this formalism, $f_{k}(t,x)$ is
interpreted as the probability amplitude for finding a particle of momentum
$k$ at the point $(t,x)$ in spacetime.  Thus, $\Phi(t,x)|0\rangle =
\Phi^{-}(t,x)|0\rangle$ represents a state of the field with a field quantum
located at $(t,x)$, and is a superposition of all one-particle states
$|1_{k}\rangle$. Similarly, the operator $\Phi^{+}(t,x)$ is interpreted as
annihilating a particle at $(t,x)$.
Thus, we may generate the Fock space using the ``position basis'' by fixing
$t$ and acting repeatedly on $\vac$ with the operators $\Phi^{-}(t,x)$.  To be
explicit when using this position basis for Fock space, we call it ``position
Fock space''.

In terms of the distributional mode operators $\phi(y)$ and $\pi(y)$, the
positive frequency field (at $t = 0$) that defines the vacuum via
(\ref{altvaccondition}) is given by
\begin{equation}
\Phi^{+}(0,y) = \frac{1}{2}\Big{(} \phi(y) + i (\Omega\star\pi)(y)\Big{)},
\end{equation}
or, in terms of the operator $\hat{\omega} := \sqrt{\partial_{y}^{2}-m^{2}}$,
\begin{equation}
 \Phi^{+}(0,y) = \frac{1}{2}\Big{(} \phi(y) + i
 (\hat{\omega}^{-1}\pi)(y)\Big{)}
 \label{posfreqphidata}
\end{equation}
where $\Omega = \widetilde{\omega_{k}^{-1}}(y)$, and $\star$ denotes
convolution.

We may view (\ref{posfreqphidata}) as defining an annihilation operator
$\Phi^{+}(0,y)$ for position Fock space by an appropriate encoding of the
field data $\phi(y)$, $\pi(y)$ into the real and imaginary parts
(respectively) of a complex quantity $\Phi^{+}(0,y)$.  ``Appropriate'' in this
case refers to equation (\ref{posfreqphidata}), which reproduces the standard
notion of positive frequency in Minkowski space.  But, as is well known from
the curved space viewpoint, there isn't a unique choice of positive frequency,
only preferred choices (even in Minkowski space!).  Therefore, we don't refer
to (\ref{posfreqphidata}) as {\it the\/} encoding.  Indeed, we now look at a
particularly convenient encoding that leads to modes called the ``pseudo plane
wave'' modes.

\subsection{Positive-Negative Frequency Splittings and Pseudo Plane Waves}

 From (\ref{posfreqphidata}), we saw that the real part of the positive
frequency field $\Phi^{+}$ (at $t = 0$) involves the data $\phi$, while the
imaginary part involves $\pi$.  We argued that we may view the forming of the
$(t = 0)$ positive frequency field as suitably encoding the (real) Cauchy data
$\phi(y)$ and $\pi(y)$ into a single complex quantity, $\Phi^{+}$ -- the
negative frequency part of the field being the complex conjugate.  This is
analogous to viewing the pair $(x,y) \mem \reals^{2}$ as real and imaginary
parts of a single complex number $z$, corresponding to the positive frequency
field.  The negative frequency field corresponds to $\overline{z}$, and the
positive-negative frequency decomposition $\Phi = \Phi^{+} + \Phi^{-}$
corresponds to the statement $2x = z + \overline{z}$.

The standard Fourier transform formalism leads to a particular encoding of the
data $\phi(y)$ and $\pi(y)$ by the plane wave expansion of $\Phi^{+}$, that of
(\ref{posfreqphidata}).  However, the simplest procedure would be to use the
above analogy with $\reals^{2}$ more directly and define
\begin{equation}
2\Phi^{+}_{ppw}(0,x) := \phi(x) + i\pi(x).   
 \label{ppwposfreq}
\end{equation}
Then, we would have $2\Phi^{-}_{ppw}(0,x) := \phi(x) - i\pi(x)$.  To see how
this might arise from a mode decomposition, consider the ``pseudo plane wave''
modes $V_{k}$ and $\overline{V}_{k}$, defined by
\begin{eqnarray}
V_{k} & := &
 \frac{1}{2\sqrt{\pi}} e^{ikx} \Big{(}\cos(\omega_{k}t) -
 \frac{i}{\omega_{k}} \sin(\omega_{k}t)\Big{)} \\
\overline{V}_{k} & = &
 \frac{1}{2\sqrt{\pi}} e^{-ikx} \Big{(}\cos(\omega_{k}t)
 + \frac{i}{\omega_{k}} \sin(\omega_{k}t)\Big{)}.
\end{eqnarray}
These modes satisfy the same normalization conditions as the plane waves:
\begin{eqnarray}
(V_{k},V_{\tilde{k}})_{KG}
 & = & \delta(k - \tilde{k}) \\
(\overline{V}_{k},\overline{V}_{\tilde{k}})_{KG}
 & = & -\delta(k - \tilde{k}) \\
(V_{k},\overline{V}_{\tilde{k}})_{KG}
 & = & 0.
\end{eqnarray}
Expanding the field in terms of the pseudo plane waves yields
\begin{equation}
 \Phi(t,x) = \int_{k} \; \Big{(}b(k) V_{k} + \overline{b}(k)
 \overline{V}(k)\Big{)} \; dk.
 \label{ppwexpansion}
\end{equation}
The Bogolubov transformation between the pseudo plane waves and distributional
modes, together with the transformation between the pseudo plane waves and the
plane waves, imply:
\begin{eqnarray}
b(k) & = & \frac{1}{\sqrt{2}} \Big{(}\tilde{\phi}(k) + i
    \tilde{\pi}(k)\Big{)} \\
\overline{b}(k) & = & \frac{1}{\sqrt{2}} \Big{(}\tilde{\phi}(-k) - i
            \tilde{\pi}(-k)\Big{)} \\
b(k) & = & \frac{1}{2\sqrt{\omega_{k}}}
         \Big{(}(1 + \omega_{k}) a(k)
         + (1 - \omega_{k}) \overline{a}(-k)\Big{)} \\
\overline{b}(k) & = & \frac{1}{2\sqrt{\omega_{k}}}
                 \Big{(}(1 - \omega_{k}) a(-k)
                 + (1 + \omega_{k}) \overline{a}(k)\Big{)}.
\label{ppwbog}
\end{eqnarray}
The first two of these equations show that the pseudo plane wave coefficients
are related to the Fourier transform of the Cauchy data $\phi(y)$ and $\pi(y)$
in essentially the same manner as in (\ref{ppwposfreq}).  Thus, for $y$-space
pseudo plane waves, one could either Fourier transform on the variable $k$, or
take the appropriate complex combination of the distributional modes to obtain
the desired expansion:
\begin{equation}
 V_{y} = \frac{1}{\sqrt{2}} (\deltafm - i \thetafm).
 \label{xspaceppw}
\end{equation}

Another advantageous property of the pseudo plane waves was noted by Dray and
Manogue in \cite{madras}.  In the case of the massless scalar field, the zero
frequency limit, $\lim_{k \rightarrow 0}U_{k},$ does not exist for the plane 
wave modes, since the normalization factor $N_{k}$ diverges.  Even if we take 
the zero frequency limit before normalizing and define
$U_{0} = \lim_{k\rightarrow0}e^{-i\omega_{k}t + ikx} = 1$,
it is real (and hence has no notion of positive frequency) and is orthogonal
to the modes $\uk{k},\ukstar{k}$, as well as to itself.  Thus, the set
$\{U_{0},\uk{k},\ukstar{k}\}$ is degenerate with respect to the Klein-Gordon
form.  It is common in practice to simply discard this mode as unphysical.
However, this is hard to justify in spacetimes with compact spatial sections,
since in this case, $U_{0}$ is a smooth function with compact support.  In
contrast, Dray and Manogue pointed out that $V_{k}$ and $\overline{V}_{k}$
{\it do\/} have well defined zero frequency limits, namely
$\frac{1}{\sqrt{2}}(1 - i)$ and $\frac{1}{\sqrt{2}}(1 + i)$.  It is also
noteworthy that the distributional modes $\thetafm$ and $\deltafm$ also
have zero norm, and that $\{U_{0},\thetafm,\deltafm\}$ is {\it not\/}
degenerate with respect to the Klein-Gordon product.  This is an explicit
example of how the distributional modes (as well as the pseudo plane wave
modes) have the ability to capture all the degrees of freedom of a system,
where the plane waves cannot.  The pseudo plane waves are one example of a
family of sets of modes we call ``generalized plane waves''.  We elaborate
further on this subject in the next section, after discussing distributional 
modes in Rindler space.

\section{Rindler Space}
\label{Rindler}

\subsection{Quantum Field Theory in Rindler Space}

The Rindler wedge \cite{BirrellandDavies,madras} is the globally hyperbolic,
open submanifold of Minkowski space corresponding to the right wedge $x>|t|$.
Canonical ``Rindler coordinates'' $(\tau,\rho)$ are defined in terms of
Minkowski coordinates as follows:
\begin{eqnarray}
 \tau & = & \frac{1}{a}\tanh^{-1}\Big( \frac{t}{x}\Big) \\
 \rho & = & \frac{1}{2a}\ln[a^{2}(x^{2} - t^{2})] \\
 t & = & \frac{e^{a\rho}}{a}\sinh(a\tau) \\
 x & = & \frac{e^{a\rho}}{a}\cosh(a\tau),
 \label{rightrindlercoords}
\end{eqnarray}
where $a > 0$ is a constant.  The line element is given by:
\begin{equation}
 ds^{2} = e^{2a\rho}(-d\tau^{2} + d\rho^{2}).
 \label{rindlerlineelement}
\end{equation}
The rays $\tau = C$, $C$ a constant, are acceptable Cauchy surfaces for the
Rindler wedge.  For definiteness we will work with the Cauchy surface
$\Sigma_R$, defined by $\tau = 0$.  The curves $\rho =D$, $D$ a constant, are
hyperbolae symmetric about the $x$-axis and have as asymptotes the lines
$x=\pm t$.  These hyperbolae represent the world lines of observers undergoing
a constant acceleration $ae^{-a\rho}$.

In the case of the massless scalar field ($m = 0$), the Klein-Gordon equation
is conformally invariant, and the line element (\ref{rindlerlineelement}) is
conformal to the Minkowski line element \cite{MDCop,BirrellandDavies}.  It
follows that the entire quantization procedure in terms of Rindler plane waves
will be formally identical to the Minkowski space quantization procedure, but
with $\rho$ replacing $x$ and $\tau$ replacing $t$.  Thus, we will have an
orthonormal basis of positive frequency solutions (with respect to Rindler
time $\tau$), $U^{R}_{l}(\tau,\rho) =
\frac{1}{\sqrt{4\nu_{l}\pi}}e^{-i\nu_{l}\tau + il\rho}$, that will split
Rindler space fields (i.e.\ solutions to the Rindler space wave equation) into
positive and negative frequency parts,
$\Phi^{R}(\tau,\rho) = {\Phi^{R}}^{+} + {\Phi^{R}}^{-}$.
In the usual way, the coefficients $a^{R}, {a^{R}}^{\dagger}$ of the field
expansion become operators satisfying the CCR (\ref{ccr}) and define a vacuum
state $\rvac$ which serves as a cyclic vector for a Fock space of states
${\cal{F}}^{R}$.

Comparing the Minkowski vacuum $\vac$ and the Rindler vacuum $\rvac$, one
finds\cite{BirrellandDavies}:
\begin{equation}
 \langle 0|{{a^{R}}^{\dagger}a^{R}}|0\rangle =
 \frac{1}{e^{\frac{2\pi\nu}{a}} - 1}.
 \label{rindlertemp}
\end{equation}
Thus, the Minkowski vacuum state contains Rindler particles (i.e.\ the notion
of particles defined by quantum field theory the Rindler space) in each mode
$\nu$, at a density ${1}/{(e^{\frac{2\pi\nu}{a}} - 1})$!  Moreover, this
spectrum of particles is precisely a Planckian spectrum at temperature
$\frac{a}{2\pi}$.  As mentioned above, curves of the form ``$\rho = D$''
represent uniformly accelerating observers, and so we may interpret the
difference in vacua by saying that uniformly accelerating observers in
Minkowski space with a quantum field in the vacuum state will ``feel
themselves immersed in a thermal bath of radiation at the temperature
$\frac{a}{2\pi}$''\cite{Unruh}.  This is the Unruh effect.

\subsection{Distributional Modes in Rindler Space}
\label{rindlerdmodes}

The Rindler space distributional and pseudo plane wave modes are {\it
formally\/} the same as the Minkowski versions, but in terms of different
variables:
\begin{eqnarray}
\rthetafm(\tau,\rho)
& = & \frac{1}{2}\Big{(}\theta(\rho - \lambda + \tau) -
 \theta(\rho - \lambda - \tau)\Big{)}  \\
\rdeltafm(\tau,\rho)
& = & \frac{1}{2}\Big{(}\delta(\rho - \lambda + \tau) +
 \delta(\rho - \lambda - \tau)\Big{)} \\
V^{R}_{l}
& = & \frac{1}{\sqrt{2}}\Big{(}\widetilde{\Theta}_{l} - i
 \widetilde{\Delta}_{l}\Big{)},
\end{eqnarray}
where $\lambda,l \mem \Sigma^{R}$.  The distributional modes $\thetafm$ are
coordinate independent, in the sense that
$\rthetafm(\tau,\rho) = \Theta_{y(\lambda)}(t(\tau,\rho),x(\tau,\rho))$.
However, the same does not hold for $\rdeltafm$ and $\deltafm$.  The modes
$\rdeltafm$ (respectively, $\deltafm$) have been defined as the $\tau$
(respectively $t$) derivative of $\rthetafm$ (respectively $\thetafm$), and
$\partial_{\tau}\neq\partial_{t}$.  This is the origin of the inequivalent
notions of positive frequency with respect to Minkowski and Rindler
coordinates in the distributional modes formalism, where it appears as
different notions of time ($t$ vs. $\tau$).

In the right Rindler wedge, we have the expansions
\begin{eqnarray}
 \Phi^{R}(\rho,\tau) & = & \int_{l} dl \Big{(}a^{R}(l) U^{R}_{l} +
 \overline{a^{R}}(l) \overline{U^{R}_{l}}\Big{)}
 \label{rindlerpwexp}  \\
 \Phi^{R}(\rho,\tau) & = & \int_{l} dl \Big{(}b^{R}(l) V^{R}_{l} +
 \overline{b^{R}}(l) \overline{V^{R}_{l}}\Big{)} \\
\Phi^{R}(\rho,\tau) & = & \int_{\lambda} d\lambda
 \Big{(} \phi^{R} (\lambda) \rdeltafm + \pi^{R} (\lambda)
 \rthetafm\Big{)}.
 \label{rindlerfmexp}
\end{eqnarray}

As in the Minkowski case, it follows from the computation of the Bogolubov
transformations that the expansion coefficients, hence the annihilation and
creation operators defined by the modes, are related as follows:
\begin{eqnarray}   
a^{R}(l)
& = & { 1 \over {2 \sqrt{{\nu}}}}
 \Big{(}{\nu}\tilde{\phi}^{R} (l) + i \tilde{\pi}^{R} (l)\Big{)} \\
\overline{a^{R}}(l)
& = & { 1 \over {2 \sqrt{{\nu}}}}
 \Big{(}{\nu} \tilde{\phi}^{R} (-l)  + i \tilde{\pi^{R} (-l)}\Big{)} \\
b^{R}(l)
& = & \frac{1}{\sqrt{2}}\Big{(}\tilde{\phi}^{R}(l) +
 i\tilde{\pi}^{R}(l)\Big{)}  \\
\overline{b}^{R}(l)
& = & \frac{1}{\sqrt{2}}\Big{(}\tilde{\phi}^{R}(-l) -
 i\tilde{\pi}^{R}(-l)\Big{)} \\
b^{R}(l)
& = & \frac{1}{2\sqrt{\nu_{l}}}\Big{(}(1 + \nu_{l})a^{R}(l)
 + (1 - \nu_{l})\overline{a}^{R}(-l)\Big{)}  \\
\overline{b}^{R}(l)
& = & \frac{1}{2\sqrt{\nu_{l}}}\Big{(}(1 - \nu_{l})a^{R}(-l)
 + (1 + \nu_{l})\overline{a}^{R}(l)\Big{)}.
\end{eqnarray}

The sets of modes $U_{l}$, $V_{l}$, and $\rthetafm,\rdeltafm$ are related by
\begin{eqnarray}
V_{l} & = &  \Big{(}\frac{1}{2} + \frac{1}{2\omega_{k}}\Big{)}U_{k} +
      \Big{(}\frac{1}{2} - \frac{1}{2\omega_{k}}\Big{)}\bar{U}_{k} \\
& = &  \frac{1}{\sqrt{2}}(\widetilde{\Theta}_{l} -
      i\widetilde{\Delta}_{l}).
\end{eqnarray}

\subsection{The Unruh Effect}
\label{unruheffect}

A standard method for deriving the Unruh and Hawking effects is to analyze the
Bogolubov transformations between the basis modes associated with the two
different coordinate systems corresponding to two sets of observers
{\boldmath${a}$} and {\boldmath${b}$}.  That one obtains a {\it thermal\/}
spectrum of ``{\boldmath${a}$} particles'' in the {\boldmath${b}$} vacuum
state follows from a particular property (say, {\sf P}) satisfied by the
Bogolubov coefficients $\boga$ and $\bogb$.  For the Unruh effect, one makes
the standard choice of basis modes in both coordinate systems -- plane waves.
Although this is a natural choice to make, the point has been stressed that
the consideration of alternatives to the plane waves may be advantageous and,
in certain spacetimes, even necessary.  Thus, the question arises if, and to
what extent, the properties of the Bogolubov coefficients that lead to the
thermal spectrum result do (or do not) depend on this choice of modes.
 
In the remainder of this section we will define a family of Bogolubov
transformations on plane waves to obtain an infinite family {\sf F} of sets of
``generalized plane wave'' modes.  We will regard each set in the family as an
alternative set of basis modes for the mode-sum formulation of quantum field
theory.  Considering both Minkowski and Rindler generalized plane waves, we
show that the Unruh effect may be derived by choosing any set of modes in the
family to serve as the basis modes, so long as corresponding sets of
generalized plane waves are used in the two different coordinate systems.
 
We give a brief review of the derivation of the Unruh effect here in 
\S\ref{unruheffect}, pointing out the property {\sf P} of the Bogolubov 
coefficients that leads to the Unruh effect. In \S\ref{genpw} and 
\S\ref{invthm} we define the family {\sf F} of sets of modes and prove the 
invariance of the property {\sf P} under changes of mode sets within 
{\sf F}.

We compute the expression for the Bogolubov coefficients $\boga$ and $\bogb$
for the Minkowski plane waves $\uk{k}$ and $U^{R}_{l}$.
\begin{eqnarray}
\boga & = & \ckgproduct{U^{R}_{l}}{\uk{k}} \\
& = & -i\int_{0}^{\infty}\, (\frac{1}{\sqrt{4 \pi \omega_{k}}})
        (\frac{1}{\sqrt{4 \pi \nu_{l}}})
        \Big{(}
        i \nu_{l} e^{ikx} e^{-a\rho(x) - il\rho(x)} +
        i \omega_{k} e^{ikx} e^{- il\rho(x)}
        \Big{)}
        \, dx \\
& = & \frac{\sqrt{\omega_{k}}}{4\pi\sqrt{\nu_{l}}}
        \int_{0}^{\infty}\,
        e^{ikx} e^{-\frac{il}{a} \ln(ax)}
        \Big{(}
        1 + \frac{\nu_{l}}{ax\omega_{k}}
        \Big{)}
        \, dx.
\end{eqnarray}
Let $x\to iy$ and rotate the resulting contour back to the
positive real axis $x>0$.  Using $\ln(iy) = \frac{i\pi}{2} + \ln(y)$,
$y \mem \reals$, we get
\begin{eqnarray}
\boga & = & \frac{i\sqrt{\omega_{k}}}{4\pi \sqrt{\nu_{l}}}
\int_{0}^{\infty}\, e^{-ky} e^{-\frac{il}{a} \ln(iay)}
             \Big{(}
             1 + \frac{\nu_{l}}{iay\omega_{k}}\Big{)}
             \, dy \\
& = & e^{\frac{l\pi}{2a}}
       \Big{[}
       \frac{i\sqrt{\omega_{k}}}{4\pi \sqrt{\nu_{l}}}
       \int_{0}^{\infty}\, e^{-ky} e^{-\frac{il}{a} \ln(ay)}
       \Big{(}
             1 + \frac{\nu_{l}}{iay\omega_{k}}
             \Big{)}\, dy
       \Big{]}.
\end{eqnarray}
By essentially the same calculation, one finds
\begin{eqnarray}
-\bogb & = & \ckgproduct{U^{R}_{l}}{\ukstar{k}} \\
& = & e^{-\frac{l\pi}{2a}}
      \Big{[}
       -\frac{i\sqrt{\omega_{k}}}{4\pi \sqrt{\nu_{l}}}
       \int_{0}^{\infty}\, e^{-ky} e^{\frac{il}{a} \ln(ay)}
       \Big{(}
             -1 + \frac{\nu_{l}}{iay\omega_{k}} \Big{)}
             \, dy
            \Big{]}.
\end{eqnarray}
Thus, we see that the Bogolubov coefficients satisfy the property
\begin{equation}
\boga = e^{\frac{l\pi}{a}} \bogbbar,
\label{property}
\end{equation}
which implies
\begin{equation}
|\boga|^{2} = e^{\frac{2l\pi}{a}} |\bogbbar|^{2}.
\end{equation}
This is the property {\sf P} from which the thermal spectrum result follows,
provided the Bogolubov identities\cite{BirrellandDavies,DM} are satisfied.

\subsection{Generalized Plane Waves}
\label{genpw}

Consider a Bogolubov transformation of the plane waves
\begin{equation}
\pmatrix{U_{k}\cr \bar{U}_{-k}\cr}\mapsto\pmatrix{V_{k}\cr\bar{V}_{-k}\cr} 
= \pmatrix{A_{k} & B_{k}\cr B_{k}  & A_{k} \cr}
\pmatrix{U_{k}\cr \bar{U}_{-k}\cr},
\end{equation}
that does not mix frequencies and is subject to
\begin{equation}
{\rm det}\pmatrix{A_{k} & B_{k}\cr B_{k} & A_{k}\cr}
= A_{k}^{2} - B_{k}^{2} = 1,
\end{equation}
so that the normalization is preserved.  Generally, we have such a
transformation for each function $\theta(k)$ if we define $A_{k} =
\cosh(\theta(k))$, $B_{k} = \sinh(\theta(k))$.  If $B_{k} \neq 0$, 
there will be mixing of positive and negative frequencies, and the 
transformed modes will define a different vacuum state than the standard 
Minkowski vacuum\cite{BirrellandDavies}.  An explicit example of such a 
transformation is given by the pseudo plane wave modes, for which
$\theta(k)=-\frac{1}{2}\ln(\omega_{k})$.  We may formally reproduce 
the above construction for plane waves in Rindler space and write the
transformation as
\begin{equation}
\pmatrix{V^{R}_{l}\cr \bar{V}^{R}_{-l}\cr} =
\pmatrix{A_{l} & B_{l}\cr B_{l} & A_{l}\cr}
\pmatrix{U^{R}_{l}\cr \bar{U}^{R}_{-l}\cr} =
\pmatrix{A_{l} U^{R}_{l} + B_{l} \bar{U}^{R}_{-l} \cr
B_{l} U^{R}_{l} + A_{l} \bar{U}^{R}_{-l}\cr}.
\end{equation}
We now have a new set of modes $\{ V_{k}, \bar{V}_{-k} \}$ in Minkowski space
and a new set $\{ V^{R}_{l}, \bar{V}^{R}_{-l} \}$ in Rindler space.

By using a different set of basis modes we obtain a different set of operator
coefficients.  It is straightforward to check that for the above class of
transformations the new operator coefficients satisfy the same canonical
commutation relations as their plane wave counterparts.  That is,
\begin{equation}
[a(k),a^{\dagger}(\tilde{k})] = \delta(k - \tilde{k}),
\label{ccr}
\end{equation}
together with
\begin{eqnarray}
b(k) & = & A_{k}\, a(k) + B_{k}\, a^{\dagger}(-k)
\label{opbog1} \\
b^{\dagger}(k) & = & A_{k}\, a^{\dagger}(k) + B_{k}\, a(-k),
\label{opbog2}
\end{eqnarray}
and $A_{k}^2 - B_{k}^2 = 1$ imply that
\begin{equation}
[b(k),b^{\dagger}(\tilde{k})] = \delta(k - \tilde{k})
\end{equation}
(with all other commutators vanishing).  Thus, the commutation relations
(\ref{ccr}) are unchanged by the transformation (\ref{opbog1}),(\ref{opbog2}).

Using (\ref{opbog1}),(\ref{opbog2}), we may compute the following vacuum
expectation values:
\begin{eqnarray}
\langle0|b^{\dagger}(k)b(\tilde{k})|0\rangle & = &
\langle0| \{A_{k}\, a^{\dagger}(k) + B_{k}\, a(-k)\}\, \{A_{\tilde{k}}\,
a(\tilde{k}) + B_{\tilde{k}}\, a^{\dagger}(-\tilde{k})\}   |0\rangle
\nonumber \\
& = & \langle0| B_{k}B_{\tilde{k}} a(-k) a^{\dagger}(-\tilde{k})
|0\rangle
\nonumber \\
& = & B_{k}B_{\tilde{k}} \langle0|  \delta(\tilde{k} - k) +
a^{\dagger}(-\tilde{k})a(-k)  |0\rangle \nonumber \\
& = & B_{k}B_{\tilde{k}} \langle0|  \delta(\tilde{k} - k) |0\rangle
+ \langle0| a^{\dagger}(-\tilde{k})a(-k)  |0\rangle \nonumber \\
& = & B_{k}^2 \, \delta(\tilde{k} - k).
\end{eqnarray}

Consider the special case where the operators $a(k),a^{\dagger}(k)$ are the
usual Minkowski plane wave operator coefficients and $b(k),b^{\dagger}(k)$ are
the pseudo plane wave operator coefficients.  The number operator for $k$-mode
particles is $b^{\dagger}(k)b(k)$.  The vacuum expectation value of the number
operator for $k$-mode particles is then infinite:
\begin{equation}
\langle0|b^{\dagger}(k)b(k)|0\rangle \rightarrow \frac{1}{4}\delta(0)
\Big{(} \frac{1}{\omega_{k}} + \omega_{k} - 2 \Big{)}.
\end{equation}

The infinity due to the delta function arises because of the infinite spatial
volume of time slices ($t = k$, $k \mem \reals$ hypersurfaces) in Minkowski
space.  In particular, if we ``put the field in a box'' (that is, assume that
the time slices are flat 3-tori instead of copies of $\reals^3$
\cite{Wald94}), the delta function becomes a Kronecker delta, and in this
case,
\begin{equation}
\langle0|b^{\dagger}(k)b(k)|0\rangle = \frac{1}{4}
\Big{(} \frac{1}{\omega_{k}} + \omega_{k} - 2 \Big{)}.
\end{equation}
The total particle content still diverges:
\begin{equation}
\sum_{k = -\infty}^{\infty}\langle0|b^{\dagger}(k)b(k)|0\rangle =
\frac{1}{4} \sum_{k = -\infty}^{\infty}
\Big{(} \frac{1}{\omega_{k}} + \omega_{k} - 2 \Big{)} = \infty.
\end{equation}
This shows explicitly that the vacuum defined by the pseudo plane wave modes
defines a Fock space that is not unitarily equivalent to the Fock space
generated by the standard Minkowski vacuum, for finiteness of
$\sum_{k=-\infty}^{\infty}\langle0|b^{\dagger}(k)b(k)|0\rangle$ is necessary
and sufficient to guarantee that the pseudo plane wave vacuum and the standard
Minkowski vacuum define unitarily equivalent Fock spaces\cite{Fulling89}.

The following section shows that, somewhat surprisingly, the modes $\{ V_{k},
\bar{V}_{-k} \}$ and $\{ V^{R}_{l}, \bar{V}^{R}_{-l} \}$ can nevertheless be
used to derive the Unruh effect.

\subsection{Invariance Theorem}
\label{invthm}

We have already noted that the if property {\sf P} is satisfied by the
Bogolubov coefficients, then the Unruh effect follows via the consistency
conditions satisfied by the Bogolubov coefficients.  We first derive some
properties of the Bogolubov transformations $\boga$ and $\bogb$ between
Minkowski and Rindler plane waves.
\begin{lemma}
The Bogolubov coefficients $\boga$ and $\bogb$
associated with the plane wave modes in Minkowski and Rindler
space satisfy
\begin{eqnarray}
 \bogabar & = & \alpha_{-l\, -k} \\
 \bogbbar & = & \beta_{-l\, -k}.
 \end{eqnarray}
\end{lemma}

Proof:

By direct computation,
\begin{eqnarray}
\bogabar & = & \overline{({U^{R}_{l}}{\uk{k})}}_{KG} \\
& = & i\int_{0}^{\infty}\,
          (\frac{1}{\sqrt{4 \pi \omega_{k}}})
          (\frac{1}{\sqrt{4 \pi \nu_{l}}})
        e^{-ikx}   e^{il\rho(x)}
        \Big{(}
        -i \nu_{l}  e^{-a\rho(x)}-
        i \omega_{k}
        \Big{)}
        \, dx \\
& = & - (\frac{1}{\sqrt{4 \pi \omega_{k}}})
        (\frac{1}{\sqrt{4 \pi \nu_{l}}})
        \int_{0}^{\infty}\,
        e^{i(l\rho(x) - kx)}
        \Big{(}
        \nu_{l}  e^{-a\rho(x)} + \omega_{k}
        \Big{)} \, dx \\
      & = & \alpha_{-l\, -k}.
 \end{eqnarray}
 
Similarly,
\begin{eqnarray}
\bogbbar & = & -\ckgproduct{U^{R}_{l}}{\bar{U}_{k}} \\
& = & -i (\frac{1}{\sqrt{4 \pi \omega_{k}}})
        (\frac{1}{\sqrt{4 \pi \nu_{l}}})
        \int_{0}^{\infty}\,
        e^{-i(l\rho(x) + kx)}
        \Big{(}
        -i\nu_{l}  e^{-a\rho(x)} - i\omega_{k}
        \Big{)} \, dx \\
        & = & - (\frac{1}{\sqrt{4 \pi \omega_{k}}})
        (\frac{1}{\sqrt{4 \pi \nu_{l}}})
        \int_{0}^{\infty}\,
        e^{i(-l\rho(x) - kx)}
        \Big{(}
        \nu_{l}  e^{-a\rho(x)} + \omega_{k}
        \Big{)} \, dx \\
        & = & \beta_{-l\, -k}.
 \end{eqnarray}

We now prove the main result.

\begin{proposition}
Property {\sf P} holds for the transformation $\{ V_{k},
\bar{V}_{-k}\} \mapsto \{ V^{R}_{l}, \bar{V}^{R}_{-l} \}$ 
if and only if {\sf P} holds for the transformation
$\{U_{k},\bar{U}_{-k}\} \mapsto \{U^R_{k},\bar{U}^R_{-k}\}$.
That is, for $\tilde{\alpha}_{lk} = \ckgproduct{V^{R}_{l}}{V_{k}}$,
$\tilde{\beta}_{lk} = -\ckgproduct{V^{R}_{l}}{\bar{V}_{k}}$,
$\alpha_{lk} = \ckgproduct{U^{R}_{l}}{U_{k}}$ and
$\beta_{lk} = -\ckgproduct{V^{R}_{l}}{\bar{V}_{k}},$
then $\tilde{\alpha}_{lk} = e^{\frac{\pi l}{a}}
\overline{\tilde{\beta}}_{lk}$
if and only if $\alpha_{lk} = e^{\frac{\pi l}{a}}
\bar{\beta}_{lk}$ .
\end{proposition}

Proof:

The Bogolubov coefficients $\tilde{\alpha}_{lk}$ may be expressed in terms of
the (untransformed) plane wave Bogolubov coefficients by substitution:
\begin{eqnarray}
\tilde{\alpha}_{lk} & = & \ckgproduct{V^{R}_{l}}{\bar{V}_{-k}} \\
& = & \ckgproduct{A_{l} U^{R}_{l} + B_{l} \bar{U}^{R}_{-l}}
    {A_{k} U_{k} + B_{k} \bar{U}_{-k}} \\
    & = & A_{l}A_{k}\ckgproduct{U^{R}_{l}}{U_{k}} +
          A_{l}B_{k}\ckgproduct{U^{R}_{l}}{\bar{U}_{-k}} + \\
& & B_{l}A_{k}\ckgproduct{\bar{U}^{R}_{-l}}{U_{k}} +
          B_{l}B_{k}\ckgproduct{\bar{U}^{R}_{-l}}{\bar{U}_{-k}}.
\end{eqnarray}
Using the lemma, we get
\begin{eqnarray}
\tilde{\alpha}_{lk}
 & = & A_{l}A_{k}\boga -
         A_{l}B_{k}\bar{\beta}_{-l\, k} +
         B_{l}A_{k}\bar{\beta}_{-l\, k} -
         B_{l}B_{k}\boga \\
 & = & \Big{(} A_{l}A_{k} -   B_{l}B_{k} \Big{)} \boga
         \Big{(} B_{l}A_{k} - A_{l}B_{k} \Big{)} \bar{\beta}_{-l\, k}.
\end{eqnarray}
Assuming $\boga = e^{\frac{l\pi}{a}} \bogbbar$, we may use this in
the above equation to obtain
\begin{equation}
\tilde{\alpha}_{lk} = \Big{(} A_{l}A_{k} -   B_{l}B_{k} \Big{)}
e^{\frac{l\pi}{a}} \bogbbar +
                  \Big{(} B_{l}A_{k} - A_{l}B_{k} \Big{)}
                  \bar{\beta}_{-l\, k}.
\label{alphaequation}
\end{equation}
Similarly, one finds
\begin{equation}
\tilde{\beta}_{lk} = \Big{(} A_{l}A_{k} -   B_{l}B_{k} \Big{)} \bogbbar +
                 \Big{(}  B_{l}A_{k} - A_{l}B_{k}\Big{)}
                   e^{-\frac{l\pi}{a}} \bar{\beta}_{-l\, k}.
\label{betaequation}
\end{equation}
By comparing (\ref{alphaequation}) and (\ref{betaequation}), we conclude that
\begin{equation}
\tilde{\alpha}_{lk} = e^{\frac{l\pi}{a}} \overline{\tilde{\beta}}_{lk},
\nonumber
\end{equation}
that is, the Bogolubov coefficients for the transformed modes possess the
\hbox{property {\sf P}.}  Applying the same argument to the inverse
transformation proves the converse.


\section{General Theory of Distributional Modes}
\label{Cauchy}

\subsection{Existence and Uniqueness of Canonical Distributional Modes}
\label{canonical}

We have so far defined distributional modes explicitly for 2-dimensional
Minkowski and Rindler spaces, and the generalization to $n$-dimensional
Minkowski and Rindler spacetimes is straightforward.  However, our desire is
to define distributional modes in general, curved spacetimes, and the key to
this issue lies in the fact that the distributional modes are closely related
to the {\it fundamental solution\/} of the Klein-Gordon equation
\cite{Dimock80,Choquet}.  This observation makes it possible to establish the
existence and uniqueness of a canonical type of distributional mode.

The family of distributional modes \{$\thetafm$\}, taken as a single
distribution, corresponds precisely to the restriction to $\Sigma$ of the
fundamental solution (the difference between the advanced and retarded Green
functions) $G(x,y)$ of the Klein-Gordon equation
\footnote{The (homogeneous) Green function $G(x,y)$ is defined (as a
distribution) on the whole of the product manifold
${\mathcal{M}}\times{\mathcal{M}}$, but in equation (\ref{thetaequalsgreen})
we are restricting $G(x,y)$ to the subset ${\mathcal{M}}\times{\Sigma}$.}:
\begin{equation}
\thetafm = G(x,y)|_{y \mem \Sigma}.
\label{thetaequalsgreen}
\end{equation}
An important result concerning the fundamental solution is that it is uniquely
defined on any globally hyperbolic spacetime\cite{Dimock80,Choquet}.  Thus,
the distributional modes \{$\Theta_y$\} have a unique, coordinate invariant
generalization to arbitrary globally hyperbolic spacetimes that requires, and
will depend on, only a choice of Cauchy surface. The distributional modes
\{$\deltafm$\} were defined as the (coordinate) time derivative of the modes
\{$\thetafm$\}, as were the Rindler versions.  As mentioned in
\S\ref{rindlerdmodes}, the Minkowski and Rindler versions of these modes
differed because of the different definitions of time coordinate.  So,
although there is a unique set of modes \{$\thetafm$\} associated to a given
Cauchy surface in a globally hyperbolic spacetime, this will not be the case
for \{$\deltafm$\}.

The Minkowski space distributional modes have the property that their nonzero
data, being delta functions, have support concentrated on a single point.  For
example, in Minkowski space, the distributional modes have the data (at $t=0$)
\begin{equation}
 \{ \thetafm , \deltafm \} = \left \{ {{0} \choose {\delta(x - y)}},
{{\delta(x - y)} \choose {0}} \right \}.
\label{minkodmodedata}
\end{equation}
It is this property that compelled their study, since it allows one to isolate
degrees of freedom associated with points in spacetime.  Thus, the
distributional mode formulation of quantum field theory would be the most
natural approach to problems of quantization in spacetimes where there are
singular points, such as the trousers spacetime.  With this in mind, we define
{\it canonical\/} distributional modes that generalize the Minkowski
distributional modes in the following sense.  For Cauchy surfaces in Minkowski
space of the form $t = C$, $C$ a constant, where $(t,\vec{x})$ are inertial
coordinates, the normal vector coincides with $\partial_{t}$.  Since every
Cauchy surface in a globally hyperbolic spacetime has a unique, future
pointing normal vector, we may generalize the Minkowski definition of
distributional modes using the normal vector to define a preferred set of
modes $\Delta_y$.
\begin{definition}\label{dmodedef}
Let $({\cal M},g)$ be a globally hyperbolic $n$-dimensional spacetime and
$\Sigma$ a Cauchy surface.  Let $x$ and $y$ be 2 independent copies of a local
coordinate for ${\cal M}$.  Restrict to $y \mem \Sigma$.  The canonical
distributional modes are defined to be the distributional solutions to the
Klein-Gordon equation, $\thetafm$ and $\deltafm$, corresponding to the data
\begin{equation}
 \{ \thetafm , \deltafm \} = \left \{ {{0} \choose {\delta(x - y)}},
{{\delta(x - y)} \choose {0}} \right\}.
\label{generaldmodedata}
\end{equation}
\end{definition}
This definition is substantiated by the following theorem:
\begin{proposition}
Let $({\cal M},g)$ be a globally hyperbolic $n$-dimensional spacetime and
$\Sigma$ a Cauchy surface.  Then there exist a unique set of distributional
modes $\{ \thetafm, \deltafm \}$.  Furthermore, the set of modes is complete
in the sense that any ${{\cal C}^{\infty}}$ solution of the Klein-Gordon
equation $(\Box - m^{2})\Phi = 0$ with compact spatial support may be uniquely
expanded in terms of the distributional modes.
\end{proposition}
Proof:

The solution to the initial value problem for the Klein-Gordon equation 
is given in terms of the homogeneous fundamental solution $G(x,y)$ by

\begin{equation}
  \iota \colon \; \data \mapsto \int_{\Sigma}\, \Big{(}\pi(y) G(x,y) -
  \phi(y) \nabla^{y}_{\hat n} G(x,y)\Big{)} \, d\Sigma^{y} = \Phi(x).
 \label{dataisomorphism}
\end{equation}
$G(x,y)$ satisfies the Klein-Gordon equation
\begin{equation}
(\Box - m^{2})G(x,y) = 0
\end{equation}
in either argument $x$ or $y$ and is antisymmetric: $G(x,y) = -G(y,x)$.
Global hyperbolicity allows us to assume without loss of generality that the
coordinates have been chosen so that $\Sigma$ corresponds to the hypersurface
$x^{0} = t = 0$ or $y^{0} = s = 0$.  It will be helpful to be explicit and
write $G(x,y) = G(t,\vec{x};s,\vec{y})$.  The equation (\ref{dataisomorphism})
then reads
\begin{equation}
\Phi(t,\vec{x}) =  \int_{\Sigma}\, \{\pi(y) G(t,\vec{x};0,\vec{y}) -
  \phi(y) \nabla^{y}_{\hat n} G(t,\vec{x};0,\vec{y})\} \, d\Sigma^{y}.
\label{datatemp}
\end{equation}
Let $\Phi(t,\vec{x})$ be the solution generated by data $\phi(\vec{y}),
\pi(\vec{y})$, where $\phi(\vec{y})$ is an arbitrary smooth function of
compact support, and $\pi(\vec{y}) = 0$.  In this case, (\ref{datatemp}) is
just
\begin{equation}
\Phi(t,\vec{x}) =  \int_{\Sigma}\, - \phi(\vec{y}) \nabla^{\vec{y}}_{\hat n}
G(t,\vec{x};0,\vec{y}) \, d\Sigma^{y}.
\label{setpitozero}
\end{equation}
Evaluating this equation at $t = 0$, we find
\begin{equation}
\phi(\vec{x}) := \Phi(0,\vec{x}) =  \int_{\Sigma}\,
- \phi(\vec{y}) \nabla^{y}_{\hat n} G(0,x;0,y)
\, d\Sigma^{y}.
\nonumber
\end{equation}
However, this is exactly the definition of the Dirac delta distribution
$\delta(\vec{x} - \vec{y})$.  Differentiate (\ref{setpitozero}):
\begin{equation}
\nabla^{x}_{\hat n}\Phi(t,\vec{x}) =  \int_{\Sigma}\,
- \phi(\vec{y}) \nabla^{x}_{\hat n} \nabla^{y}_{\hat n}
G(t,\vec{x};0,\vec{y}) \, d\Sigma^{y}.
\nonumber
\end{equation}
Then,
\begin{equation}
0 = \pi(\vec{x}) =  \nabla^{x}_{\hat n}
\Phi(0,\vec{x}) = -\int_{\Sigma}\, \phi(\vec{y})
\nabla^{x}_{\hat n}
\nabla^{y}_{\hat n} G(0,\vec{x};0,\vec{y})d\Sigma^{y}.
\nonumber
\end{equation}
As $\phi(\vec{y})$ is arbitrary, we must have
$\nabla^{x}_{\hat n}\nabla^{y}_{\hat n} G(0,\vec{x};0,\vec{y}) = 0$.
Alternatively, let $\Phi(t,\vec{x})$ be the solution generated by data
$\phi(\vec{y}), \pi(\vec{y})$, where $\pi(\vec{y})$ is an arbitrary smooth
function of compact support, and $\phi(\vec{y}) = 0$.  Then (\ref{datatemp})
reduces to
\begin{equation}
\Phi(t,\vec{x}) =  \int_{\Sigma}\, \pi(y) G(t,\vec{x};0,\vec{y})  \,
d\Sigma^{y}.
\nonumber
\end{equation}
Evaluate this equation at $t = 0$:
\begin{equation}
0  = \phi(\vec{x}) = \Phi(0,\vec{x}) =  \int_{\Sigma}\, \pi(y)
G(0,\vec{x};0,\vec{y})  \, d\Sigma^{y}.
\nonumber
\end{equation}
As $\pi(\vec{y})$ is arbitrary, we must have $G(0,\vec{x};0,\vec{y}) = 0$.  
Lastly, evaluating the normal derivative of $\Phi$ on $\Sigma$ yields
\begin{equation}
\pi(\vec{x}) =  \nabla^{x}_{\hat n} \Phi(0,\vec{x}) = 
\int_{\Sigma}\, \pi(\vec{y}) \nabla^{x}_{\hat n} G(0,x;0,y) \, d\Sigma^{y}.
\nonumber
\end{equation}
Collecting these results we conclude that for each fixed $\vec{y}\mem\Sigma$,
$G(t,\vec{x};0,\vec{y})$ is the unique distributional solution to the
Klein-Gordon equation having data ${{0} \choose {\delta(\vec{x} - \vec{y})}}$,
and $\nabla^{y}_{\hat n} G(t,\vec{x};0,\vec{y})$ is the unique distributional
solution to the Klein-Gordon equation having data ${-\delta(\vec{x} - \vec{y})
\choose {0}}$.  Define $\thetafm$ and $\deltafm$ as in
(\ref{generaldmodedata}).  The completeness condition then follows from
(\ref{dataisomorphism}).

Thus, we may define a quantization procedure for any globally hyperbolic
spacetime by first choosing a Cauchy surface and using
(\ref{generaldmodedata}) to expand the field in terms of distributional modes.
A choice of positive-negative frequency would be made by defining the positive
frequency field operator
\begin{equation}
2\Phi^{+}_{ppw}(0,x) := \phi(x) + i\pi(x).   
\end{equation}
The position Fock space would then be generated by the vacuum state, defined by
\begin{equation}
 \Phi^{+}(t,\vec{x})\vac = 0 \; \; \; \; \forall (t,\vec{x}) \mem\Sigma.
\end{equation} 
One may alternatively choose different encodings of the expansion coefficients
into the positive frequency operator to define different vacua as desired.
For example, in the case of a static spacetime with compact spatial sections,
one may Fourier decompose with respect to Killing time and use the encoding as
in (\ref{dmvaccondition}) to obtain a vacuum analogous to the Minkowski plane
wave vacuum.

Canonical distributional modes were defined above for any choice of Cauchy
surface in any globally hyperbolic spacetime, and a prescription for second
quantization was given.  In Section \ref{Rindler}, distributional modes in
Rindler space were constructed on the $\tau = 0$ hypersurface, but this
construction is not a special case of the canonical construction.  This is due
to the fact that the unit normal vector field does not coincide with the
timelike vector field corresponding to the (Rindler) coordinate time
derivative used to construct the family $\{\Delta_{\lambda}\}$.  This
situation can be dealt with by relaxing the requirement that the unit normal
derivative be used in the definition of $\Delta_{\lambda}$.  Instead,
associate distributional modes with a choice of Cauchy surface {\it and\/} a
choice of timelike vector field $\xi$, defining the modes $\Delta_{\lambda}$
as $\xi (\Theta_{\lambda})$.  For static spacetimes such as the Rindler
spacetime, a natural choice of timelike vector field to use would be the
Killing vector field.  However, as far as the mathematical formalism is
concerned, any timelike vector field can be used.  Equation
(\ref{dataisomorphism}) would then be used to determine the expansion
coefficients.  Therefore, once a Cauchy surface and timelike derivative is
chosen, and the distributional mode expansion made, second quantization may
proceed as in the canonical case.

\section{DISCUSSION}
\label{Discussion}

In Minkowski space, our distributional modes correspond more naturally to
pseudo plane waves rather than plane waves, and in particular are {\it not\/}
Poincar\'e invariant.  Nevertheless, we showed that pseudo plane waves can be
used in place of plane waves to deduce important physical properties, such as
the Rindler temperature.  The fact that our distributional modes are related
to the fundamental solution, which leads to our existence and uniqueness
results, further suggests that pseudo plane waves may deserve further study in
their own right.

At first sight, the distributional modes have a simple definition in terms of
data on an initial data surface.  In Rindler space, however, the definition of
the distributional modes depended in a crucial way on Rindler coordinates.
Although it turns out that the modes $\rthetafm$ are invariantly defined on
any globally hyperbolic spacetime, the modes $\rdeltafm$ were defined as the
derivatives of $\rthetafm$ with respect to the Rindler time coordinate, and
therefore are not equivalent to their Minkowski counterparts.  In this
framework, therefore, the fundamental role of a family of observers appears as
the choice of time coordinate used to differentiate the $\Theta$ modes to
obtain the $\Delta$ modes.

Nonetheless, from the results of the theory of PDEs, we were able to produce a
canonical generalization of the distributional modes.  This construction most
directly generalizes the Minkowski case in that the modes are defined using a
Cauchy surface and the unit normal derivative, and then prescribing delta
function data as in (\ref{generaldmodedata}).  Work is in progress on the
generalization of the distributional modes for a given coordinate system and
Cauchy surface.

A procedure based on canonical distributional modes for the second
quantization of the scalar field in an arbitrary globally hyperbolic spacetime
is readily generalized from the Minkowski procedure.  There is no a priori
need for Fourier transformation nor the need to find solutions explicitly
other than the distributional mode solutions.  Moreover, the canonical
distributional mode solutions exist in any globally hyperbolic spacetime and
have the same simple expression in terms of Cauchy data. To demonstrate
the utility of this formalism, one should apply it to specific models, such as
the trousers spacetime.  Also desirable would be a rigorous treatment of the
functional analysis involved in the formalism, as well as a detailed
examination of how the formalism fits in with the many other approaches to
curved space quantum field theory.  These issues are all being actively
pursued.

\newpage
\section*{Acknowledgments}

We owe a considerable debt to Corinne A. Manogue for inspiring this research
with her idea of using the distributional modes {\it as modes}.  A.A. would
like to thank Robert Wald and Robert Geroch for helpful discussions and the
Enrico Fermi Institute at the University of Chicago for hospitality.  This
work forms part of a dissertation\cite{dissertation} submitted to Oregon State
University (by A.A.)\ in partial fulfillment of the requirements for the
Ph.D.\ degree in mathematics, and was partially supported by NSF Graduate
Research Fellowship 9616167-DGE.

\end{document}